\documentstyle[prb,aps,epsf,psfig,multicol]{revtex}
\newcommand{\la}{\langle}
\newcommand{\ra}{\rangle}
\newcommand{\ua}{\uparrow}
\newcommand{\da}{\downarrow}
\newcommand{\rar}{\rightarrow}
\begin{document}
\draft

\title{Local Moment Formation
in the Periodic Anderson Model with Superconducting Correlations}

\author{M. A. N. Ara\'ujo and N. M. R. Peres}
\address{Departamento de F\'{\i}sica, Universidade de \'Evora,
Rua Rom\~ao Ramalho, 59, P-7000-671 \'Evora Codex, Portugal\\
Centro de F\'{\i}sica da Universidade do Minho, Campus Gualtar, P-4700-320,
Portugal}

\author{P. D. Sacramento}
\address{Departamento de F\'{\i}sica and  CFIF, Instituto Superior 
T\'ecnico, Av. Rovisco Pais, 1049-001 Lisboa, Portugal}

\date{\today}

\maketitle

\begin{abstract}
We study  local moment formation
in the presence of superconducting correlations among the $f-$electrons
in the periodic Anderson model.
Local moments form if  the 
Coulomb interaction $U>U_{cr}$.
We find that $U_{cr}$ is
considerably stronger in the presence of superconducting correlations
than in the non-superconducting system. 
Our study is done for various values of the $f-$level energy and 
electronic density.
The smallest critical
$U_{cr}$ values occur for the case where the number of $f-$ 
electrons per site is equal to one. In the presence of  
$d-$wave superconducting correlations we find that local moment 
formation presents a quantum phase transition as function of pressure. 
This quantum phase transition separates a region where local moments and 
$d-$wave superconductivity coexist from another region characterized by
a superconducting ground state with no local moments. We discuss the possible
relevance of these results to experimental studies of the competition between
magnetic order and superconductivity in $\rm CeCu_2Si_2$.
\end{abstract}
\vspace{0.3cm}
\pacs{PACS numbers: 75.20.Hr, 71.27.+a, 74.70.Tx}
\begin{multicols}{2}

The superconducting and magnetic properties of heavy-fermion materials
have attracted much attention
mainly due to their non-conventional
character. \cite{varma85,estrela00} All  these materials have very
large specific heat coefficients $\gamma$, indicating very large effective
masses, hence the designation  {\it heavy fermions}.
The complexity 
of these systems arises from the interplay between Kondo screening of local
moments, the antiferromagnetic (RKKY) interaction between the moments and 
superconducting correlations between the heavy quasi-particles. On the 
experimental side,  these systems exhibit phases in which  antiferromagnetic
ordering of the local moments may coexist 
with unconventional  superconductivity, and/or phases with 
no magnetic ordering. 
The latter could be either  due to Kondo screening of the local moments or to 
a spin liquid  arrangement of the local moments. 

For systems that exhibit both superconductivity
and antiferromagnetism, usually $U$-based heavy-fermions,
the ratio between the N\'eel temperature $T_N$
and the superconducting critical temperature $T_c$ is of the order of
$T_N/T_c\sim 1-10$, with  coexistence of both types of order below $T_c$.
The coexistence of both types of order can be tuned  by external
parameters such as external pressure 
or changes in the stoichiometry.\cite{lonzarich98,ishida99}
Examples of  heavy-fermion materials 
which exhibit antiferromagnetic and superconducting order 
at low temperature  are  $URu_2Si_2$ and  $U_{0.97}Th_{0.03}Be_{13}$. 
It has recently been found that $UPd_2Al_3$ 
($T_N=14.3$ K and $T_c=2$ K) and $UNi_2Al_3$  ($T_N=4.5$ K and $T_c=1.2$ K)
show coexistence of superconductivity
and local moment antiferromagnetism. 
\cite{lonzarich98,steglich93,feyerherm94,bernhoeft98,also00}
In the $Ce$-based heavy-fermions however typically magnetism competes
with superconductivity.
In the prototype heavy-fermion system Ce$_x$Cu$_2$Si$_2$ the competition 
of $d-$wave superconductivity  and magnetic order 
was clearly identified in a small range of $x$ values 
around $x\simeq 0.99$.\cite{ishida99}

A description of the normal state  properties of the heavy-fermion
systems has been attempted assuming a generalization of the impurity
Anderson model to the lattice. \cite{newns87,Millis87}
In the Anderson lattice the energy of a single 
electron in an $f-$orbital (e. g. $4f^1$)
is $\epsilon_0$, and the energy of two electrons 
in the same $f-$orbital ($4f^2$) is $2\epsilon_0+U$, where $U$ is 
the on-site Coulomb repulsion. The energy of the  $4f^2$ state  is 
much larger than the energy of the $4f^1$ state. 
In the case of interest the empty $f-$orbital lies below the
Fermi level $\epsilon_F$, whereas the doubly occupied $f-$orbital
has higher energy than $\epsilon_F$. Since in many systems
$\epsilon_0+U\gg \epsilon_F$ the simplifying limit $U\rar \infty$ 
is taken is some theoretical studies.

The limit $U=\infty$ has been studied using the slave boson 
technique.\cite{Millis87,coleman84,houghton88,peres00} 
In particular, it has been 
shown that superconducting instabilities arise in the $p$ 
and $d$-wave channels because
of the effective (RKKY) interaction between the 
$f-$electrons.\cite{houghton88,lavagna87}
Mean-field studies of superconductivity in the Anderson 
lattice both at finite $U$ 
and at $U=\infty$ have recently been done.\cite{peres00,robaszkiewicz87}
Nevertheless, the coexistence of superconducting correlations 
and magnetic ordering
of local moments in heavy-fermion systems  has not yet, to our knowledge, 
been studied theoretically.

In this work we aim to establish the conditions for  the appearance
of local moments  in the  Anderson lattice when 
superconducting correlations among the $f-$electrons are also present.
In order to decide whether a $f$-site behaves as a local moment we use
a criterion identical to that introduced by  Anderson for the single
 impurity problem\cite{anderson}, which we shall now briefly review.
 The procedure in Ref \cite{anderson} consists of making 
 a Hartree-Fock decoupling of   the interaction at the impurity ($f$ site):
\begin{equation}
U \hat n_{\uparrow}^f  \hat n_{\downarrow}^f \rightarrow U \la  
\hat n_{\uparrow}^f\ra 
 \hat n_{\downarrow}^f+
U  \hat n_{\uparrow}^f \la  \hat n_{\downarrow}^f\ra 
- U \la  \hat n_{\uparrow}^f\ra \la  \hat n_{\downarrow}^f\ra \,, 
\label{andersondecoupling}
\end{equation}
where $ \hat n_{\uparrow}^f,  \hat n_{\downarrow}^f$ 
are the number operators at the $f$ site and 
their expectation values $\la \hat n_{\uparrow}^f\ra, 
\la \hat n_{\downarrow}^f\ra$ must be found 
self-consistently. Then, the  symmetry beaking  solutions  with 
$
m=\la \hat n_{\uparrow}^f\ra - \la \hat n_{\downarrow}^f\ra \neq 0
$
correspond to the  local moment regime. Therefore, the local moments arise
as the  symmetry-breaking local minima,
 with $ \la \hat n_{\uparrow}^f\ra \neq \la \hat n_{\downarrow}^f\ra$,
 of the effective action for the impurity in a Hartree-Fock  decoupling scheme.
Such a mean-field treatment (\ref{andersondecoupling}) 
does not account for  the dynamics of the local
moment 
which arises from its effective interaction with 
the conduction electrons.

In this work we propose to extend such ideas to the lattice of $f$-sites
taking phenomenologically into account the presence of superconducting correlations 
between the $f$-electrons.  
 Because we consider  a lattice of $f$ sites, we have to look for mean-field
solutions  with some  previously chosen spatial 
arrangement of the moments which 
will be taken as either ferromagnetic or antiferromagnetic. 
Since heavy fermion systems tend to  exhibit antiferromagnetic order,
we have chosen this type of broken symmetry
state. Some comments are also made on the
ferrromagnetic case. We recall that 
this mean-field treatment leaves out  the dynamics of the local
moments. Such dynamics lies beyond the scope of this work but would 
possibly allow to describe the nature of the magnetic phases of
 the system including  Kondo screening, or ordering 
due to  RKKY interaction, or a spin liquid  arrangement 
of the local moments, etc. 

Our study of the interplay between superconducting correlations
and local moment formation may be relevant to the understanding of
 recent studies of $\rm CeCu_2Si_2$  samples near stoichiometric composition, 
where it has been observed that a $d$-wave superconducting phase expels 
a magnetic ``A phase'' under increasing pressure \cite{gegenwart,bruls}.

We wish to investigate the appearance of local moments in
the Anderson lattice in the presence of superconducting correlations.
Because a microscopic description of superconductivity 
in heavy-fermion systems is still lacking we opt to treat superconductivity 
by explicitly adding a phenomenological pairing term $H_J$ to the Anderson lattice
Hamiltonian  for spin 1/2 electrons:
\begin{equation}
H=H_0+H_U + H_J\,,
\label{hi}	
\end{equation}
where
\begin{eqnarray}
H_0&=&\sum_{\vec k,\sigma}(\epsilon_{\vec k}-\mu)
c_{\vec k,\sigma}^{\dag}c_{\vec k,\sigma}\ 
+\sum_{i,\sigma}(\epsilon_0-\mu)f_{i,\sigma}^{\dag}f_{i,\sigma} \nonumber\\
&+&V\sum_{i,\sigma}\left(c_{i,\sigma}^{\dag}f_{i,\sigma}
+f_{i,\sigma}^{\dag}c_{i,\sigma}\right)
\label{hcf}\,,\\
H_U&=&U\sum_{i}\hat n_{i,\uparrow} \hat n_{i,\downarrow}\,,\\
H_J&=&J\sum_{\vec{k}}
\left(
f^{\dagger}_{\vec{k},\ua} 
{f^{\dagger}}_{-\vec{k},\da}\Delta(\vec k)
+\,{\rm H.c.}\right) \label{hj}\,,
\end{eqnarray}
where $\epsilon_{\vec k}$ is the dispersion of the $c-$electrons,
$\epsilon_0$ is the bare energy of the 
localized $f-$states, $V$
is the hybridization matrix element (assumed $\vec k$ independent),
$U$ is the on-site Coulomb interaction, 
and $ \hat n_{i,\sigma}=f_{i,\sigma}^{\dag}f_{i,\sigma}$.
The gap function 
$\Delta(\vec k)$ can be written as 
$\eta(\vec k)\Delta$, where $\eta(\vec k)$ indicates the 
pairing  symmetry and $
\Delta=\frac 1 {N_s}\sum_{\vec k}
\la f_{-\vec k,\da}f_{\vec k,\ua} \ra\eta_{\vec k}$.
To simplify the calculations we consider a two
dimensional system. The results for three dimensions are
similar. In 2D we have$
\eta_{\vec{k}}^{(s)}= 
\cos(k_x) + \cos(k_y)$,
$\eta_{\vec{k}}^{(p,i)} =  \sqrt{2}\ \sin(k_i)$,
$\eta^{(d)}_{\vec{k}} = \cos(k_x) - \cos(k_y)$,
for $s$, $p$, and $d$ wave, respectively. 

The local moment formation is investigated by applying the decoupling 
(\ref{andersondecoupling}) to the term $H_U$. We seek mean-field solutions
such that the  electron occupation at the $f$-site $i$ is given by
\begin{equation}
\la f^{\dag}_{i,\sigma}f_{i,\sigma}\ra=$ $\frac{1}{2}n_f+$
$\frac{1}{2}\sigma m
\cos(\vec R_i\cdot \vec Q),
\label{ff}
\end{equation}
where $n_f$ and   $\vec R_i$ denote the mean 
number of $f$ electrons and
the position of the $i$ lattice site, respectively, and $\sigma=\pm 1$. This 
choice corresponds to a spatially periodic 
arrangement of the local moments where 
$\vec{Q}$ is the ordering vector.
Since we shall perform the calculation on a 2D square lattice
the choice $\vec{Q}=0$ gives ferromagnetic order 
and the choice $\vec{Q}=(\pi , \pi)$ gives antiferromagnetic order.
 
The mean-field parameters $m$, $n_f=n_{\ua}^f+n_{\da}^f$,
and $\Delta$ are obtained from the minimization of the
free energy. 
The chemical potential is determined imposing 
total particle number conservation $n=n_c+n_f$.
We perform the calculations for  a  
conduction band dispersion of  form: 
$
\epsilon_{\vec k}=-2t\sum_{i=x,y} \cos(k_i)\,.
$
All the energies
and temperatures in this work are measured in units of the hopping
integral $t$, and we have set $t=1$.
The possible  pairing symmetries 
have been studied separately.

\begin{figure}
\begin{center}
\epsfxsize=7cm
\epsfbox{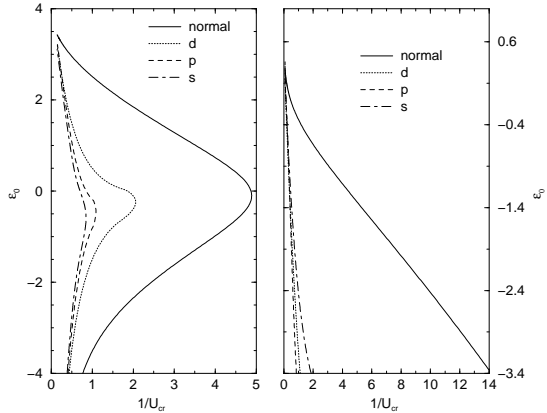}
\end{center}
\caption{Relation between the  critical $U_{cr}$ 
and the $f$ level bare energy 
$\epsilon_0$, for the parameters  $J=-1.7$, $V=0.4$, $t=1$,  
and: {\bf left panel:} $n=2$;  {\bf right panel:} $n=1$.  
The  curve for the normal system corresponds 
to $J=0$. Calculations were performed
for temperature $T=10^{-3}$.}
\label{fig1}
\end{figure}

For a given set of  parameters $n$, $\epsilon_0$, $V$, and $J$ one may ask
what is the minimum value of the local repulsion, $U_{cr}$, for which 
ground-state solutions with $(\Delta,m) \neq 0$ appear. 
Figure  \ref{fig1} shows 
$1/U_{cr}$ for each  $\epsilon_0$ and two
electronic densities. The local moment regime  appears 
for $U>U_{cr}$.
For comparison we also show $U_{cr}$ for a non-superconducting system ($J=0$).
It is seen from Figure \ref{fig1} that the appearance of local moments in
a superconducting system requires  considerably stronger $U$ than in
a  normal system. 
For each curve the  $f$ level occupancy $n_f$  increases upon increasing 
 $\epsilon_0$ and the values of $\epsilon_0$ at which $U_{cr}$ is minimum
(left panel of Figure  \ref{fig1})  correspond to $n_f=1$.
 Therefore we also see that  smaller $U_{cr}$ occurs
when $f$ level occupancy is close to one. 
If $n_f<1$ then $U$ can be increased without destroying superconductivity but
if $n_f>1$ then increasing $U$ will eventually destroy superconductivity.
\cite{peres00,robaszkiewicz87}
We note that 
the relative position of the superconducting curves in the two panels
of Figure \ref{fig1}
is different.  

We have also considered the case of 
a ferromagnetic arrangement of the local moments ($Q=0$).
Our calculations show that, for the same parameters, 
ferromagnetic solutions require much stronger $U$ 
values
($1/U_{cr}\leq 0.16$)  even in the absence of superconductivity.
This seems to be an indication that, in a lattice, there is a tendency for 
local moments to form in an antiferromagnetic arrangement.

The dependence of
the N\'eel  and  superconducting temperatures on pressure has been measured 
in some heavy-fermion systems.\cite{lonzarich98} In those studies
the N\'eel temperature is found to decrease as the applied  pressure increases 
and  superconducting  order is found to develop in a limited range of applied
pressures when the N\'eel temperature is reduced below $\sim$1K. 

\begin{figure}
\begin{center}
\epsfxsize=7cm
\epsfbox{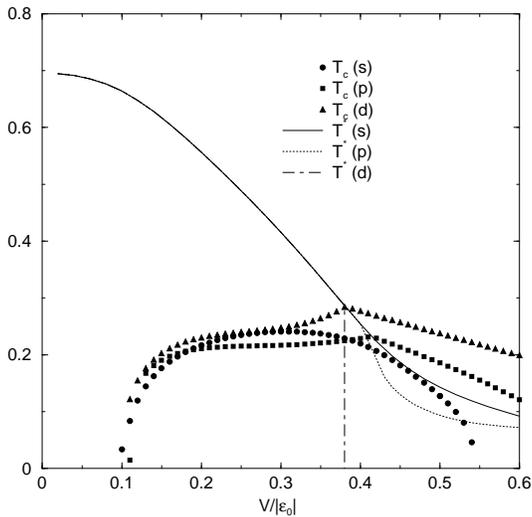}
\end{center}
\caption{Mean field magnetic ($T^{*}$) and superconducting ($T_c$)
critical  temperatures  versus  $V/\vert\epsilon_0\vert$,
 for a constant ratio $V/t=0.5$. The parameters are $U=2.8$, $J=-1.7$, 
$\epsilon_0=-3.4$, and $n=1$. The various curves for the magnetic
transition coincide up to $V/\epsilon_0 \sim 0.38$.}
\label{fig3}
\end{figure}

It is interesting to see how the mean field critical temperatures in our model
vary with
the model parameters which, in principle, should depend on externally appplied
pressure. Increasing  pressure should, presumably, make
the hybridization $V$ and the conduction band hopping $t$ increase  while
keeping $U$ and possibly $J$ almost constant.\cite{lacroix99}
In Figure \ref{fig3} we plot the mean field critical temperatures as
function of $V$ while keeping 
the ratio $V/t$ fixed. An interesting feature  occurs
for the $d$-wave pairing symmetry: if the temperature
$T^{*}$ that marks the onset of $m\neq 0$ solutions is higher than the 
superconducting temperature $T_c$ then
$d$-wave superconductivity and local moments coexist at  low temperature; but
once $T_c$ becomes greater than $T^{*}$ the latter drops abruptly and we cannot
find magnetic solutions at the lowest temperatures.
At zero temperature we have 
a quantum phase transition that can  be tuned using
the external pressure: as pressure is reduced (at zero temperature) the
ground state nature of the system changes  from non-magnetic but 
superconducting to magnetic and superconducting at
the critical value $V_c/\epsilon_0 \sim 0.38$. 
 We find this
transition to be of first order.
Such  behaviour is not observed for $s-$ or $p-$wave symmetries.

Going beyond mean field, and
in general terms, the local moments due to the $f$-electrons
are progressively quenched as the temperature lowers. In
dilute systems the picture is well understood as due to the
Kondo screening by the conduction electrons. In dense
systems however the picture is more involved. At low
temperatures the local moments are not completely quenched.
For several $U$-based materials like $URu_2Si_2$, $UPt_3$ or
$UPd_3$ the remaining moments are quite small of the order of
$0.01-0.03 \mu_B$ but for other systems like $UPd_2Al_3$ the
local moment is quite large of the order of $0.85 \mu_B$. A
reason is the Kondo-Nozi\`{e}res compensation theorem that
states that in the lattice there are not enough available
conduction electrons to quench the local moments.

The $Ce$-based compounds like $CeCu_2Si_2$ and $CeCu_6Au$
show competition between magnetism and superconductivity
if doped and/or under pressure with a phase higher in
temperature where spin fluctuations are significant. The behavior
of these systems has been interpreted as due to the vicinity
to a quantum critical point \cite{SteglichJ}. Two pictures
arise however \cite{ColemanCe}: in the first one the Kondo
temperature is high (the moments are quenched at a finite temperature)
and when the system approaches the quantum critical point there are
no free moments (assuming that quenching is complete). Then the
system has to order due to a Fermi surface instability of the spin density
wave type. Another possible situation is one in which the moments are
not completely quenched down to $T=0$ and are free to orient themselves
leading to magnetism. In the case of $CeCu_{6-x}Au_x$ it has been
recently found that the second picture holds \cite{ColemanCe}. On the
other hand the high value of the Kondo temperature for the $CeCu_2Si_2$
compound \cite{ishida99} indicates possibly that the first scenario
should hold. Our results show that $d$-wave pairing excludes magnetism
through a quantum phase transition. The treatment of a spin-density-wave
would be mathematically similar and also lead to the exclusion of the 
spin-density-wave by $d$-wave pairing  as  pressure increases.
This means that our results might be of relevance for understanding the 
interplay between superconductivity and 
magnetic order in the $d$-wave heavy-fermion compound $\rm CeCu_2Si_2$.
This system exhibits a magnetic ``A phase'' at low temperature 
whose detailed nature is not yet known. Increasing pressure reduces
the critical temperature $T_A$ of the A phase. 
Recent studies\cite{ishida99,gegenwart,bruls,luke}
of    $\rm CeCu_2Si_2$  samples near stoichiometric composition 
have shown that a $d$-wave superconducting phase expels  the magnetic
``A phase'' when $T_A$ approaches $T_c$
under increasing pressure.  Our results in Figure \ref{fig3}
show that $d-$wave superconductivity destroys the local moments  when 
$T^*$ meets $T_c$ as function of increasing hybridization.

In summary,
we have studied the formation of local 
moments in the periodic Anderson model in the presence of superconducting
correlations. We identify the local moment regime with the symmetry-breaking
saddle-point of the effective action for the $f$-sites in a Hartree-Fock
decoupling of local Coulomb repulsion. We have found that local moments 
tend to order  antiferromagnetically. We note that this 
does not imply that antiferromagnetic order arises in a
Slater scenario. Magnetic order in heavy fermion systems is actually 
believed to be due to the effective  RKKY interaction between local moments.
This means that once we know the region in parameter space where local
moments form, further studies of magnetic order, Kondo screening, etc,  
possibly  require an  effective interaction Hamiltonian for the 
local moments.\cite{new}
By simulating the effect of increasing pressure, we have found that 
d-wave superconductivity competes (rather than coexists) with local moment 
formation  at low temperature above some critical pressure.


This research was supported by Portuguese program PRAXIS 
XXI under grant number
2/2.1/FIS/302/94.

\end{multicols}

\end{document}